\documentstyle[prl,aps,multicol,epsfig]{revtex}
\draft
\title{Anomaly in the stability limit of liquid helium~3} 
\author{Fr{\'e}d{\'e}ric Caupin and S{\'e}bastien Balibar}
\address{Laboratoire de Physique Statistique de l'Ecole Normale Sup\'erieure \\
 associ\'e aux Universit\'es Paris 6 et Paris 7 et au CNRS \\
 24 rue Lhomond 75231 Paris Cedex 05, France}
\author{Humphrey J. Maris}
\address{Department of Physics, Brown University, Providence, Rhode Island 02912}
\date{13 September 2001}

\newcommand{\mrm}[1]{{\rm{#1}}}
\newcommand{\Eb}{E_\mrm{b}}
\newcommand{\beq}{\begin{equation}}
\newcommand{\eeq}{\end{equation}}
\newcommand{\Rc}{R_\mrm{c}}
\newcommand{\kb}{k_\mrm{B}}
\newcommand{\Pcav}{P_\mrm{cav}}
\newcommand{\Ps}{P_\mrm{s}}
\newcommand{\cs}{c_\mrm{S}}
\newcommand{\ct}{c_\mrm{T}}

\newcommand{\etal}{\textit{et~al.}}

\begin{document}
\maketitle
\begin{abstract}
We propose that the liquid-gas spinodal line of $^3\mrm{He}$ reaches a minimum at $0.4\,\mrm{K}$. This feature is supported by our cavitation measurements. We also show that it is consistent with extrapolations of sound velocity measurements. Speedy [J. Phys. Chem. {\bf 86}, 3002 (1982)] previously proposed this peculiar behavior for the spinodal of water and related it to a change in sign of the expansion coefficient $\alpha$, i. e. a line of density maxima. $^3\mrm{He}$ exhibits such a line at positive pressure. We consider its extrapolation to negative pressure. Our discussion raises fundamental questions about the sign of $\alpha$ in a Fermi liquid along its spinodal.
\end{abstract}
\begin{multicols}{2}

Because of its high purity, liquid helium is an ideal material in which to study homogeneous nucleation. During the past years, a powerful experimental method has been developed to investigate the gas nucleation in the stretched liquid. It consists in focusing a high amplitude sound wave in a small region far from any wall, thus making heterogeneous nucleation unlikely. In this Letter, we present a detailed analysis of results recently obtained in $^3\mrm{He}$~\cite{Caupin01}. We show that these results disagree to some extent with existing theoretical descriptions of the liquid at negative pressure. We then propose an estimation of its stability limit (the spinodal line, where $(\partial V/\partial P)_\mrm{T}$ and $(\partial V/\partial T)_\mrm{P}$ diverge) based on sound velocity measurements by Roach~\etal~\cite{Roach83}. The spinodal $\Ps(T)$ we obtain exhibits a minimum at $0.4\,\mrm{K}$ and gives a temperature dependence of the cavitation pressure consistent with our measurements. For water, Speedy~\cite{Speedy82} previously proposed a spinodal with a minimum and showed that this change in slope of $\Ps(T)$ was linked to the change in sign of the isobaric expansion coefficient $\alpha$. This change in sign, which corresponds to a line of density maxima (LDM), also occurs in liquid $^3\mrm{He}$ at positive pressure. We give an extrapolation of the LDM at negative pressure and finally give theoretical arguments about the sign of $\alpha$ near the spinodal line.

In any substance below its saturated vapor pressure, the liquid phase can be metastable since an energy barrier $\Eb$ must be overcome for liquid-gas separation to occur, that is for a bubble to nucleate in the liquid. The experiments in liquid helium reported in Ref.~\cite{Caupin01} measure the probability of these cavitation events. For a given experiment performed in an experimental volume $V$ and during an experimental time $\tau$, this probability is, at a pressure $P$ and a temperature $T$:
\beq
\Sigma(P,T)=1-\exp\left[-\Gamma_0 V \tau \exp\left( - \frac{\Eb(P,T)}{\kb T}\right) \right]\; ,
\label{eq:Scurve}
\eeq
where $\Gamma_0$ is a prefactor discussed below. The experimental measurements are reproduced by the asymmetric S-curve formula of Eq.~(\ref{eq:Scurve}) with great accuracy~\cite{Caupin98}. $\Gamma_0$ has the dimensions of frequency times an inverse volume. It is natural to estimate $\Gamma_0$ as an attempt frequency $\nu$ at which the fluctuations try to overcome the nucleation barrier multiplied by the density of the critical nuclei which can be taken to be spheres of radius $\Rc$~\cite{Xiong89,Guilleumas93,Pettersen94}. Typically, $\Rc$ is around $1 \,\mrm{nm}$ and the attempt frequency varies from $\kb T/h$ to $\Eb/h$; all the different estimates thus lie between $5\times 10^{36} \,T$ and $1.5\times 10^{38} \,T\,\mrm{m^{-3}\,s^{-1}\,K^{-1}}$. Pettersen~{\etal}~\cite{Pettersen94} have calculated $V$ and $\tau$ for the experimental method which uses an acoustic wave to produce a negative pressure swing in the liquid. For $^3\mrm{He}$ and for a $1\,\mrm{MHz}$ acoustic wave as in Ref.~\cite{Caupin01}, this gives $V \tau = 1.2\times 10^{-22} \,\mrm{m^3\,s}$~\cite{Vtau}. The theoretical estimates of the factor $\Gamma_0 V \tau$ thus vary from $6\times 10^{14} \,T$ to $1.8\times 10^{16} \,T\,\mrm{K^{-1}}$. Although this range extends over two orders of magnitude, it does not significantly affect the value of the energy barrier: for $\Sigma=0.5$, all estimates give $\Eb=(34 \pm 3) \kb T$.

\begin{figure}
\begin{center}
\psfig{file=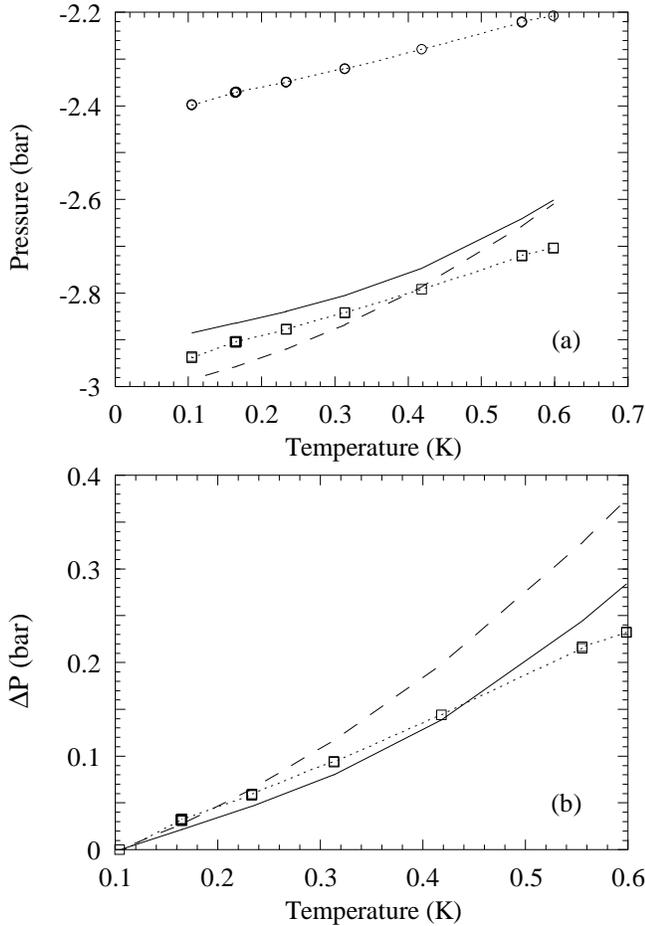,width=8.5cm}
\end{center}
\caption{(a) Comparison between experimental and theoretical cavitation pressures. The experimental limits for the cavitation line are given by circles for the upper bound and squares for the lower one; dotted lines are guides to the eye. Other lines are theoretical cavitation lines calculated with $\Gamma_0 V \tau= 6\times 10^{14} \,T\,\mrm{K^{-1}}$ using two different sources for the spinodal pressure (see Fig.~\ref{fig:comparPs}): Guilleumas~{\etal}~\protect\cite{Guilleumas93} (dashed line) and this work (solid line). (b) Temperature variation $\Delta P=\Pcav(T)-\Pcav(0.1\,\mrm{K})$ of the cavitation pressure. $\Delta P$ obtained from the experimental lower bound is given by squares; the dotted line is a guide to the eye. Other lines are calculated $\Delta P$ according to Guilleumas~{\etal}~\protect\cite{Guilleumas93} (dashed line) and this work (solid line).}
\label{fig:comparPexp}
\end{figure}

Recently, Caupin and Balibar~\cite{Caupin01} have given experimental limits for the pressure at which the cavitation probability is one half (the cavitation line $\Pcav(T)$) in liquid $^3\mrm{He}$ and liquid $^4\mrm{He}$. The upper and lower bounds deduced from their measurements in $^3\mrm{He}$ are shown in Fig.~\ref{fig:comparPexp}(a). The actual cavitation line is located between these limits and parallel to them. For both isotopes, Caupin and Balibar made a comparison with the spinodal pressure at which the nucleation barrier vanishes; they used the values obtained by Maris at low temperature: $\Ps=-9.6 \,\mrm{bar}$ for $^4\mrm{He}$~\cite{Maris94} and $\Ps=-3.15 \,\mrm{bar}$ for $^3\mrm{He}$~\cite{Maris95}. In fact, because of thermal or quantum
{\noindent}fluctuations in the liquid, the cavitation pressure is always higher than the spinodal pressure, and the difference can be calculated from Eq.~(\ref{eq:Scurve}) if the expression of $\Eb(P,T)$ is known.

Maris~\cite{Maris95} has calculated $\Eb(P)$ at low temperature by a density functional method; close to the spinodal, his results are well represented by a power law:
\beq
\frac{\Eb}{\kb} = \beta (P-\Ps)^{\delta}\; ,
\label{eq:Eb}
\eeq
with $\beta=47.13\,\mrm{K\,bar^{-3/4}}$ and $\delta=3/4$. However, to calculate the cavitation pressure up to $0.6\,\mrm{K}$, we need to know the temperature dependence of $\Eb$. The strongest
\begin{figure}
\begin{center}
\psfig{file=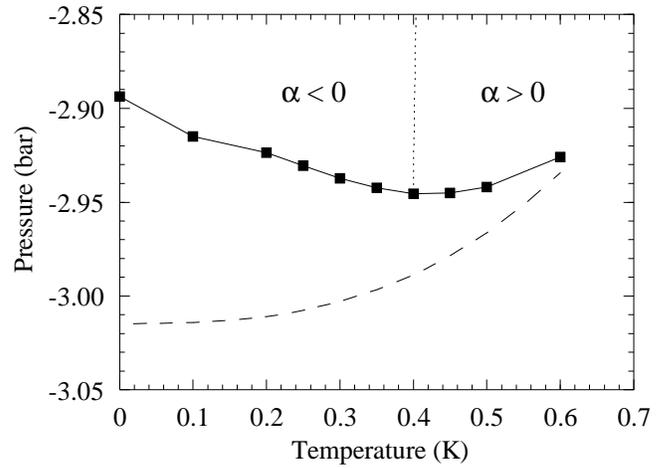,width=8.5cm}
\end{center}
\caption{Comparison between two theoretical estimates of the spinodal line: Guilleumas~{\etal}~\protect\cite{Guilleumas93} (dashed line) and this work (full squares; the solid line is a guide to the eye). The spinodal found in this work shows a minimum at $0.4\,\mrm{K}$. The dotted line is a linear extrapolation of the LDM as measured by Boghosian~{\etal}~\protect\cite{Boghosian66} between $0$ and $11\,\mrm{bar}$ (see Fig.~\ref{fig:alphaexp}). Notice that the pressure scale is different from Fig.~\ref{fig:comparPexp}.}
\label{fig:comparPs}
\end{figure}
{\noindent}source of this dependence is that the spinodal pressure varies with temperature; therefore we write $\Eb(P,T) = \Eb(P-\Ps(T))$ and assume that Eq.~(\ref{eq:Eb}) remains valid at higher temperature with parameters $\beta$ and $\delta$ held constant. The temperature dependence of the cavitation pressure follows from Eq.~(\ref{eq:Scurve}):
\beq
\Pcav(T) = \Ps(T) + \left[ \frac{T}{\beta} \ln\left(\frac{\Gamma_0 V \tau}{\ln 2} \right) \right]^{1/\delta}\; .
\label{eq:Pcav}
\eeq

Guilleumas~{\etal}~\cite{Guilleumas93} have calculated the spinodal pressure as a function of temperature; they find a monotonically increasing pressure as shown on Fig.~\ref{fig:comparPs}. Guilleumas~{\etal} have also calculated $\Eb(P,T)$ and deduced the cavitation line $\Pcav(T)$. Their estimate of $\Pcav(T)$ does not agree with our results. However, this estimate was based on a value of $V \tau$  of $2.5\times 10^{-19}\,\mrm{m^3\,s}$ which is much larger than the one corresponding to our experiment. We have therefore used our approximation of $\Eb(P,T)$ and their result for $\Ps(T)$ to calculate $\Pcav(T)$ for the appropriate value of $V \tau$. Fig.~\ref{fig:comparPexp}(a) shows the line obtained with Eq.~(\ref{eq:Pcav}) for the lowest possible value of the prefactor, namely $\Gamma_0 V \tau = 6\times 10^{14} \,T\,\mrm{K^{-1}}$. We first note that this line does not lie between the experimental limits at low temperature~\cite{otherPs}; however, this could be due to some systematic error in the lower bound estimate, which would result in shifting each pressure by the same amount (up to $\pm 0.15\,\mrm{bar}$ as explained in Ref.~\cite{Caupin01}). Let us focus on the temperature dependence of the cavitation pressure, which is free of this systematic error: Fig.~\ref{fig:comparPexp}(b) displays the quantity $\Delta P=\Pcav(T)-\Pcav(0.1\,\mrm{K})$ for the three lower lines of Fig.~\ref{fig:comparPexp}(a). The temperature variation of $\Delta P$ we obtain with the $\Ps(T)$ curve of Guilleumas~{\etal} is stronger than the experimental one.

How can we explain this discrepancy? Of course one can assume that the theory fails in estimating the value of $\Gamma_0$. However, to reproduce the experimental temperature dependence of $\Pcav$ would require $\Gamma_0 V \tau$ to be at least $3$ orders of magnitude smaller than expected. We do not see any reasons to support this hypothesis. Instead, we think that the experimental measurements question the shape of the spinodal limit.

Before proceeding further, we need to recall how the spinodal pressure $\Ps$ can be obtained: Maris' method~\cite{Maris94,Maris95} consists in extrapolating measurements of the sound velocity $c$ at positive pressure with a law of the form $c = \left[b \, (P - \Ps)\right]^{1/3}$.
Maris used for $c$ the measurements of Abraham~{\etal} at low temperature~\cite{Abraham72}. We used the same method with a set of data from Roach~{\etal}~\cite{Roach83}: they measured the first sound velocity along isochores at starting pressures from $1.6$ to $28.1\,\mrm{bar}$ and as a function of temperature from $0.01$ to $0.6\,\mrm{K}$. The spinodal line we obtained is shown in Fig.~\ref{fig:comparPs}: the spinodal pressure reaches a minimum of $-2.9\,\mrm{bar}$ around $T=0.4\,\mrm{K}$. We would like to emphasize that none of the previous estimates of the spinodal pressure in liquid $^3\mrm{He}$~\cite{Guilleumas93,Maris95,Solis92,Casulleras00} has mentioned the possible existence of a minimum in the spinodal line. The new shape of the spinodal curve we propose is sufficient to remove the discrepancy stated above: using again Eq.~(\ref{eq:Pcav}) with the value $\Gamma_0 V \tau = 6\times 10^{14} \,T\,\mrm{K^{-1}}$, we find a cavitation line which has a temperature dependence consistent with the experimental results (see Fig.~\ref{fig:comparPexp}(b)). One can wonder if the use of Maris' method to estimate the spinodal pressure is relevant for the points at higher temperature. Thermodynamically, the spinodal line is the locus of points at which the isothermal sound velocity $\ct$ vanishes, whereas the measured sound velocity is the adiabatic one, $\cs$. This was stated before~\cite{Balibar98}, along with the fact that the difference vanishes at zero temperature. We made the appropriate corrections. Both spinodal lines, obtained either with $\cs$ or $\ct$, show a minimum, and their difference is less than $30\,\mrm{mbar}$. We have actually plotted the spinodal obtained with the isothermal data in Fig.~\ref{fig:comparPs}, and we used it in the above reasoning.

We shall now turn to the physical origin of such a minimum in the spinodal. A similar behavior was first proposed by Speedy in the case of water~\cite{Speedy82}. This was the basis of the stability limit conjecture introduced to explain anomalies of supercooled water: in his theory, the liquid-gas spinodal was assumed to be reentrant at temperatures below $35$\textsuperscript{o}C. A review of this topic also describing alternative theories can be found in Ref.~\cite{DebenedettiBook}. Following a thermodynamical analysis first developed in
the case of helium~\cite{Ahlers76}, Speedy shows that close to the spinodal the sign of the isobaric thermal expansion coefficient $\alpha$ of the liquid is the same as the sign of ${\rm d}\Ps/{\rm d}T$.
\begin{figure}
\begin{center}
\psfig{file=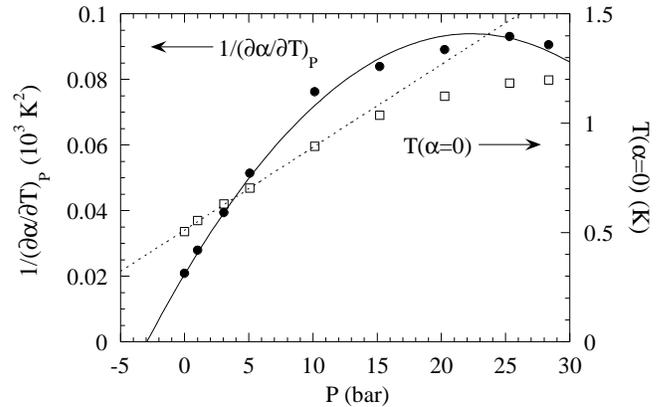,width=8.5cm}
\end{center}
\caption{Temperature of density maximum (empty squares) and inverse of $(\partial \alpha/\partial T)_\mrm{P}$ around $\alpha=0$ (full circles) as functions of pressure, derived from measurements by Boghosian~{\etal}~\protect\cite{Boghosian66}. The dotted line shows the extrapolation of the LDM used in Fig.~\ref{fig:comparPs}. The solid line is a parabolic fit to $1/(\partial \alpha/\partial T)_\mrm{P}$ forced to vanish at the pressure of the minimum in the spinodal.}
\label{fig:alphaexp}
\end{figure}
{\noindent}Therefore, if the locus of points such that $\alpha(P,T)=0$ intersects the spinodal, this results in an extremum in the curve $\Ps(T)$. Water and $^3\mrm{He}$ have in common that both liquids exhibit a LDM: in some temperature range, they expand upon cooling. Therefore they may exhibit such a minimum in the spinodal. We have tried to adapt Speedy's conjecture to the case of $^3\mrm{He}$. The measurements of Roach~{\etal} give the expansion coefficient, but unfortunately they are made in a region of the phase diagram where $\alpha$ is always negative; to obtain the LDM in $^3\mrm{He}$, we need to know the temperature where $\alpha$ vanishes. Therefore we used measurements by Boghosian~{\etal}~\cite{Boghosian66}, which extend to higher temperatures and agree well with Roach values in the region where both sets overlap; the result is shown in Fig.~\ref{fig:alphaexp}. A simple linear extrapolation of the LDM for pressures below $11\,\mrm{bar}$ extends down to the minimum in the spinodal as shown in Fig.~\ref{fig:comparPs}. In his original paper~\cite{Speedy82}, Speedy shows that the expansion coefficient at the spinodal undergoes a jump from $-\infty$ to $+\infty$ at the temperature at which the LDM meets the spinodal. To find some evidence to support this prediction, we follow Speedy's analysis for water and derive the slope $(\partial \alpha/\partial T)_\mrm{P}$ around $\alpha=0$ for each isobar in the measurements by Boghosian~{\etal} This slope should diverge when the pressure reaches the spinodal. Fig.~\ref{fig:alphaexp} shows that the experimental values are consistent with this prediction.

We now give some theoretical arguments concerning the sign of $\alpha$. The negative sign of $\alpha$ in $^3\mrm{He}$ at low temperature was first observed experimentally in 1958 by Lee~{\etal}~\cite{Lee58}. The same year, Brueckner and Atkins~\cite{Brueckner58} pointed out how this behavior was related to the variation of the effective mass with density.
Indeed, using a Maxwell relation, we can write:
\beq
\alpha = - \frac{1}{V} \left(\frac{\partial S}{\partial P}\right)_T\; .
\label{eq:alpha(S)}
\eeq
In the Fermi liquid region, the heat capacity $C_\mrm{V}$ is linear in $T$ and we have $S=C_\mrm{V}=(m^\ast/m)\,C_\mrm{F}$ where $C_\mrm{F}$ is the heat capacity of the Fermi gas. Using Greywall's measurements of the effective mass~\cite{Greywall86} and extrapolating them at negative pressure as we did before~\cite{Balibar98}, we find that $\alpha$ given by Eq.~(\ref{eq:alpha(S)}) remains negative down to the spinodal. Of course we should consider the corrections to the linear regime of the heat capacity and their evolution close to the spinodal. We see two sources of corrections. The first one is the contribution of phonons to the heat capacity, which varies as $(T/c)^3$, where $c$ is the sound velocity; this term could become important near the spinodal where the isothermal sound velocity vanishes. However, this is relevant only for the long wavelength phonons: as stated by Lifshitz and Kagan~\cite{Lifshitz72}, the first correction to the linear dispersion gives for small momentum:
\beq
{\omega_k}^2=k^2(c^2+2\rho\lambda k^2)\; ,
\eeq
where $\lambda$ is a constant. As the spinodal is approched, the dispersion relation thus becomes quadratic. A calculation shows that the correction to $\alpha$ remains negligible at temperatures of interest here. We also note that, if the sound remains adiabatic at small $k$ close to the spinodal, the use of $\cs$, which does not vanish at $\Ps$, instead of $\ct$ would further reduce the phonon contribution. The second correction is due to the coupling of the quasiparticles to the incoherent spin fluctuations and varies as $T^3 \ln T$. This effect has been studied by Greywall~\cite{Greywall83}, who has shown that its amplitude decreases when pressure decreases; it is not clear to us if this is the case until the spinodal is reached, and this point requires further investigation. 

We will end with some remarks about $^4\mrm{He}$. Its liquid phase exhibits two lines of density extrema at positive pressure. Clearly, it would be interesting to know how they extend into the metastable liquid region to determine the shape of the spinodal line. This problem is also related to the behaviour of the roton minimum in the excitation dispersion curve and to the slope of the superfluid transition line in $(T,\!P)$ coordinates at negative pressures. Several authors~\cite{He4} have addressed some of these issues, but a unified picture is still lacking.

In this Letter, we have studied the temperature dependence of the cavitation pressure in liquid $^3\mrm{He}$. We have shown that recent measurements disagree with existing theories. We then proposed a new picture for liquid $^3\mrm{He}$ at negative pressure. From the pressure and temperature dependence of the sound velocity in $^3\mrm{He}$, we obtained a liquid-gas spinodal different from what was previously predicted: this new spinodal is reentrant, that is to say that the curve $\Ps(T)$ exhibits a minimum of $-2.9\,\mrm{bar}$ at $T=0.4\,\mrm{K}$. This new feature in the phase diagram of liquid $^3\mrm{He}$ agrees with our measurements of the temperature dependence of the cavitation pressure. Following an analysis by Speedy in the case of water, we have emphasized the relationship between this behavior and the negative expansion coefficient in $^3\mrm{He}$. Finally we have given theoretical arguments to estimate this expansion coefficient at negative pressures.

\end{multicols}

\begin{references}

\bibitem{Caupin01} F.~Caupin and S.~Balibar, Phys. Rev. B {\bf 64}, 064507 (2001).

\bibitem{Roach83} P.~R.~Roach, Y.~Eckstein, M.~W.~Meisel and L.~Aniola-Jedrzejek, J.~Low~Temp.~Phys. {\bf 52}, 433 (1983).

\bibitem{Speedy82} R.~J. Speedy, J.~Phys.~Chem {\bf 86}, 3002 (1982).

\bibitem{Caupin98} F.~Caupin, P.~Roche, S.~Marchand, and S.~Balibar, J. Low Temp. Phys. {\bf 113}, 473 (1998).

\bibitem{Xiong89} Q.~Xiong and H.~J.~Maris, J.~Low Temp. Phys. {\bf 77}, 347 (1989).

\bibitem{Guilleumas93} M.~Guilleumas, M.~Pi, M.~Barranco, J.~Navarro, and M.~A.~Solis, Phys. Rev. B {\bf 47}, 9116 (1993).

\bibitem{Pettersen94} M.~S.~Pettersen, S.~Balibar and H.~J.~Maris, Phys. Rev.~B {\bf 49}, 12062, (1994).

\bibitem{Vtau} The expression of $V \tau$ obtained in Ref.~\cite{Pettersen94} actually depends on the minimum pressure reached, but we have checked that this does not affect our conclusions.

\bibitem{Maris94} H.~J.~Maris, J.~Low Temp. Phys. {\bf 94}, 125 (1994).

\bibitem{Maris95} H.~J.~Maris, J. Low Temp. Phys. {\bf 98}, 403 (1995).

\bibitem{Boghosian66} C.~Boghosian, H.~Meyer and J.~E.~Rives, Phys.~Rev. {\bf 146}, 110 (1966).

\bibitem{otherPs} All other theoretical estimates of $\Ps(T=0)$ are more negative than the one given in Ref.~\cite{Guilleumas93}, and using them to calculate $\Pcav(T=0)$ would increase this discrepancy.

\bibitem{Abraham72} B.~M.~Abraham, D.~Chung, Y.~Eckstein, J.~B.~Ketterson, and P.~R.~Roach, J.~Low~Temp.~Phys. {\bf 6}, 521 (1972).

\bibitem{Solis92} M.~A.~Sol{\'\i}s and J.~Navarro, Phys. Rev. B. {\bf 45}, 13080 (1992).

\bibitem{Casulleras00} J.~Casulleras and J.~Boronat, Phys. Rev. Lett. {\bf 84}, 3121 (2000).

\bibitem{Balibar98} S.~Balibar, F.~Caupin, P.~Roche, and H.~J.~Maris, J. Low Temp. Phys. {\bf 113}, 459 (1998). 

\bibitem{DebenedettiBook} P.~G.~Debenedetti, {\it Metastable liquids} (Princeton University Press, Princeton, 1996).

\bibitem{Ahlers76} G.~Ahlers, in {\it The physics of liquid and solid helium}, edited by K.~H.~Bennemann and J.~B.~Ketterson (Wiley, New York 1976), Chap. 2.

\bibitem{Lee58} D.~M.~Lee, J.~D.~Reppy, and H.~A.~Fairbank, Bull. Am. Phys. Soc. Ser. II. {\bf 3}, 339 (1958)

\bibitem{Brueckner58} K.~A.~Brueckner and K.~R.~Atkins, Phys. Rev. Lett. {\bf 1}, 315 (1958).

\bibitem{Greywall86} D.~S.~Greywall, Phys. Rev. B. {\bf 33}, 7520 (1986).

\bibitem{Lifshitz72} I.~M.~Lifshitz and Yu.~Kagan, Zh. Eksp. Teor. Fiz. {\bf 62}, 385 (1972) [Sov. Phys. JETP {\bf 35}, 206 (1972)]. See also Ref.~\cite{Xiong89}.

\bibitem{Greywall83} D.~S.~Greywall, Phys. Rev. B. {\bf 27}, 2747 (1983).

\bibitem{He4} C.~E.~Campbell, R.~Folk, and E.~Krotscheck, J. Low Temp. Phys. {\bf 105}, 13 (1996); S.~C.~Hall and H.~J.~Maris, J. Low Temp. Phys. {\bf 107}, 263 (1997); S.~M.~Apenko, Phys. Rev. B. {\bf 60}, 3052 (1999); G.~H.~Bauer, D.~M.~Ceperley, and N.~Goldenfeld, Phys. Rev. B. {\bf 61}, 9055 (2000); V.~P.~Skripov, Usp. Fiz. Nauk, {\bf 170}, 559 (2000) [Phys. Usp. {\bf 43}, 515 (2000)].

\end{references}
\end{document}